\documentclass[11pt,twoside]{article}
\usepackage{asp2004}
\usepackage{psfig}
\usepackage{epsf}
\usepackage{graphics}
\usepackage{lscape}
\begin{document}
\title{The Supernovae Associated with Gamma-Ray Bursts}
\author{Thomas Matheson}
\affil{Harvard-Smithsonian Center for Astrophysics, 60 Garden Street,
Cambridge, MA  02138}

\begin{abstract}
Supernovae (SNe) were long suspected as possible progenitors of
gamma-ray bursts (GRBs).  The arguments relied on circumstantial
evidence.  Several recent GRBs, notably GRB~030329, have provided
direct, spectroscopic evidence that SNe and GRBs are related.  The SNe
associated with GRBs are all of Type Ic, implying a compact
progenitor, which has implications for GRB models.  Other peculiar
Type Ic SNe may help to expand understanding of the mechanisms
involved.
\end{abstract}

\section{Introduction}
The mechanisms that produce gamma-ray bursts (GRBs) have long been
among the mysteries of modern astrophysics.  There have been a wide
variety of models (see M{\'e}sz{\'a}ros 2002 for a review of the
theories of GRBs).  The discovery of optical afterglows opened a new
window on the field (see, e.g., van Paradijs, Kouveliotou, \& Wijers
2000).  Early identification of the optical afterglows led to the
measurement of redshifts for several GRBs (e.g.,~GRB~970508: Metzger
et~al.~1997), providing definitive evidence for their cosmological
origin.  Observations at other wavelengths, especially radio, have
revealed many more details about the bursts (e.g., Berger et al. 2000;
Frail et al. 2003).  This only applies to the long-duration bursts at
this point.

Models that use supernovae (SNe) to explain GRBs have been part of the
discussion from the start (e.g., Colgate 1968; Woosley 1993; MacFadyen
\& Woosley 1999).  Strong evidence for the GRB-SN association was
first provided by GRB~980425; the optical afterglow was not found, but
an unusual Type Ic SN was seen in the error box of the GRB (Galama et
al. 1998; Patat \& Piemonte 1998).  Many other circumstantial
arguments were made to connect GRBs with SNe.  Direct evidence, where
an afterglow is seen to transform into an SN, was first seen with
GRB~030329 (Stanek et~al. 2003; Matheson et~al. 2003; Hjorth
et~al. 2003; Kawabata et~al. 2003; Kosugi et~al. 2004).  Subsequently,
two other cases have been identified: GRB~021211 (Della Valle
et~al. 2003) and GRB~031203 (Malesani et~al. 2004).  In this paper, I
will review the evidence for the GRB-SN association and describe the
nature of the SNe that are associated with GRBs.

\section{The Circumstantial Case}

There were many indirect pieces of evidence that showed an association
between GRBs and SNe.  Most of these actually related GRBs to massive
stars, but, since massive stars do undergo core collapse to become
SNe, the connection was implicit.  On the large scale, the host
galaxies of GRBs showed strong star formation, indicating that massive
stars were present (e.g., Hogg \& Fruchter 1999).  More detailed
studies found that the locations of GRBs within their host galaxies
were associated with the regions that contained massive stars
(e.g.,~Bloom, Kulkarni, \& Djorgovski 2002).

Other circumstantial arguments derived from evidence related to the
interaction of the GRB with its local environment.  Massive stars in
general have stellar winds, and the observations of the afterglows are
consistent with models of interaction with these winds (e.g.,
Chevalier \& Li 2000).  A few GRB afterglows have shown strong
absorption features at relatively high velocities (a few thousand km
s$^{-1}$), the best case being GRB~021004.  For that GRB, these
features were interpreted as mass-loss shells from a Wolf-Rayet
progenitor, again implying a massive star (Mirabal et~al. 2003;
Schaefer et~al. 2003).  Another claim of evidence for association with
SNe came from reported detections of line features in the X-ray
afterglows of some GRBs (e.g., Piro et~al. 1999; Reeves et~al. 2002)
that would result from material synthesized in the collapse, followed
by a GRB after a period of time.  Recent reanalysis of all
observations of X-ray afterglows, however, indicates that these
earlier claims are not statistically significant (Sako, Harrison, \&
Rutledge 2004).

Perhaps the strongest element in the circumstantial case for the
GRB-SN association was the presence of bumps in the afterglows of many
GRBs.  The traditional optical afterglow of a GRB decays as a power-law.
Late-time deviations from this power law were consistent in their
timing and brightness with an SN having exploded at the same time as
the GRB (e.g., Bloom et~al. 1999).  Rebrightening, though, is not
necessarily direct evidence of a supernova.  That would require a
spectroscopic detection.

\section{The Early Clues}

The first evidence for an SN associated with a GRB came when SN~1998bw
was found in the error box of GRB~980425 (Galama et~al. 1998).  The SN
was of Type Ic (see Filippenko 1997 for a review of SN types), but it
was peculiar, with evidence for unusually high velocities (Patat et
al. 2001).  This event was odd in many ways.  The GRB itself was
relatively weak (Woosley, Eastman, \& Schmidt 1999) and the radio
emission from the SN was extraordinary, both in luminosity and the
rapidity of its appearance (Kulkarni et al. 1998).  This was clearly
an object of great interest, but it was not conclusive proof that SNe
and GRBs are related.  It is, though, the framework upon which all
further claims of GRB-SN connections are built.

A weaker case is GRB~011121 (Garnavich et~al. 2003a; Bloom et
al. 2002).  A late-time bump in the light curve was observed from the
ground and with \emph{HST}.  At $z = 0.36$, the supernova component
would have been relatively bright.  A spectrum of the GRB taken during
the rebrightening does not show any obvious features of an SN, but the
color evolution derived from broad-band photometry is consistent with
a supernova (designated SN~2001ke; Garnavich et~al. 2003a).

\section{The Definitive Case}

While SN~1998bw was a strong hint of the GRB-SN association, no
traditional optical afterglow was seen.  Without that direct
association, the link between GRBs and SNe was still in question.  
The `monster burst' of 2003, GRB~030329 was to provide that link.  The
burst was unusually bright in gamma-rays, implying that it was
relatively close.  Spectroscopy of the afterglow soon confirmed a
low redshift of 0.1685 (Greiner et~al. 2003).  As the afterglow faded,
subtle features appeared in the normally flat power-law spectra of
the afterglow.  By subtracting a continuum based upon the early shape
of the spectrum, this structure was revealed as the spectrum of an
unusual Type Ic SN similar to SN~1998bw, designated SN~2003dh
(Garnavich et~al. 2003b).  Within a few days, the SN became the
dominant component in the spectrum (Stanek et~al. 2003; Hjorth et
al. 2003; Kawabata et~al. 2003).

\begin{figure}[!ht]
\plotfiddle{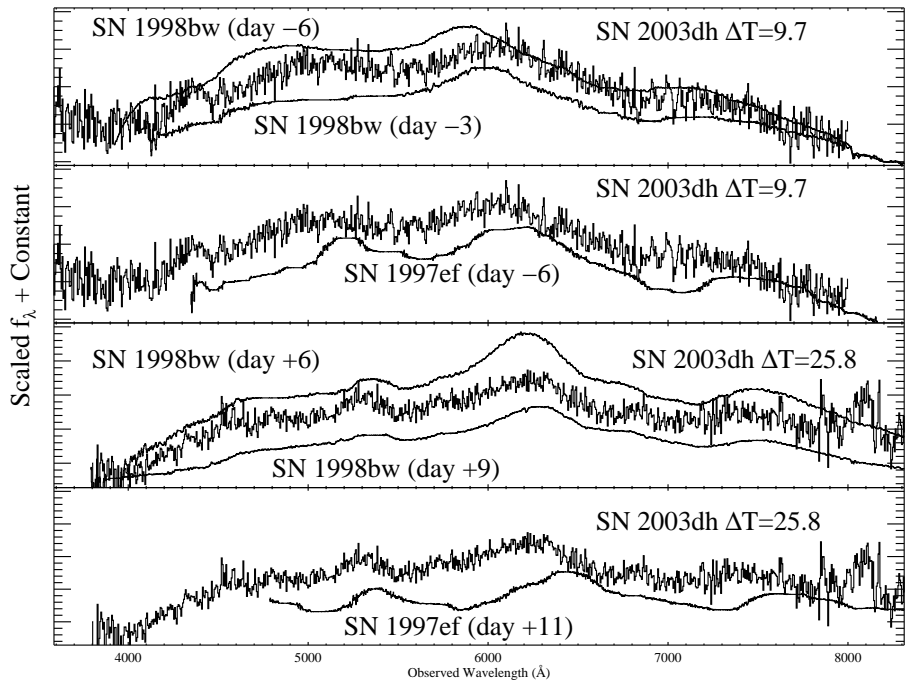}{3.5in}{90}{50}{50}{190}{-40}
\caption{Spectra of SN~2003dh after the continuum of the GRB afterglow
has been removed.  The $\Delta$T value is the time since the GRB
(observed frame).  In all cases, SN~1998bw is the best match, day -6
for $\Delta$T~=~9.7 days and day +6 for $\Delta$T~=~25.8 days.  Note
also that between $\Delta$T~=~9.7 days and $\Delta$T~=~25.8 days is
13.8 rest-frame days, consistent with the evolution of SN~1998bw.  The
time of $\Delta$T = 0 days would correspond with 14 days before
maximum for SN~1998bw, a reasonable rise time for a Type Ic SN.  See
Matheson et~al. (2003) for more details.}
\end{figure}

Using the early power-law continuum spectrum as a model, one could
decompose the observed spectra at later times into two separate
components: GRB afterglow and SN spectrum.  Using a least-squares
technique, the best match for the SN among the low-redshift sample was
SN~1998bw (Matheson et~al. 2003).  In fact, taking into account
cosmological time dilation, the spectroscopic evolution of SN~2003dh
almost exactly matched SN~1998bw (Figure 1).  Models of these spectra
are presented by Mazzali et al. (2003; see also Mazzali's contribution
in this volume).

An important point about the appearance of the SN was that the light
curve did not show the bump that is supposed to be the characteristic
of a rebrightening caused by the SN (see Matheson et~al. 2003 and
Lipkin et~al. 2004 for a discussion of the light curve).  Without the
spectroscopic confirmation, the SN in GRB~030329 would still be a
matter of contention.

Nebular-phase spectra of SN~2003dh show a spectrum similar to a
typical Type Ic SN.  Kosugi et~al. (2004) present a spectrum at an age
of $\sim$3 months.  A spectrum obtained with the Keck telescope by
Filippenko, Chornock, \& Foley (2004) in December of 2003 is much like
a normal Type Ic SN (Bersier et~al., in preparation).

\section{Other Supernovae Associated with Gamma-Ray Bursts}

Following the discovery of SN~2003dh, reexamination of spectra of an
earlier burst yielded some evidence for a SN component.  Della Valle
et~al. (2003) found that a spectrum of the afterglow of GRB~021211 had
structure similar to an SN.  In this case, though, the SN did not
match SN~1998bw or any other peculiar Type Ic SN.  Rather, it was most
similar to SN~1994I, a relatively normal Type Ic\footnote{It should be
pointed out that ``normal'' is a fairly subjective term to be applied
to Type Ic SNe.  They are a very heterogeneous group, and it is not
clear that one can define a truly normal Type Ic SN.  See Matheson et
al. (2001), and references therein, for a more complete discussion of
the nature of Type Ic SNe.}.  

A more clear example came with GRB~031203.  Despite high foreground
reddening, spectroscopy with the VLT revealed an SN component,
designated SN~2003lw (Malesani et~al. 2004).  For this SN, SN~1998bw
was again a good match.  Of the four SNe with clear GRB associations,
three show remarkably similar spectra.

\section{Why is Type Ic Significant?}

The classification scheme for SNe (see Filippenko 1997)
is based upon spectroscopic features.  The Type II SNe show hydrogen,
while those of Type I do not.  Type Ia SNe have a distinctive silicon
feature and show an elemental pattern consistent with the theory that
they arise from thermonuclear disruption of a white dwarf.  The other
Type I SNe (i.e., without the strong silicon absorption) are
distinguished either by the presence (Ib) or absence (Ic) of helium.

A strong circumstantial case has tied Types II, Ib, and Ic to the same
underlying mechanism of core collapse.  A transition between Type II
and something like a Ib was observed for SN~1993J (e.g., Filippenko,
Matheson, \& Ho 1993), strengthening this connection and providing an
explanation for the subclasses.  Models of the progenitor of SN~1993J
postulated that a massive star in a binary had lost most of its
hydrogen layer, leaving a small amount that produced the early
spectrum of a Type II, and then revealing the helium layer below (see
Matheson et~al. 2000 for a summary and an extensive list of
references).  

Study of a large number of Type Ib/c SNe indicated that the SNe Ic had
less massive envelopes than the SNe Ib (Matheson et~al. 2001).  This
implied that the Ib SNe have lost their hydrogen, leaving the helium
layer, while the SNe Ic have lost hydrogen and helium, leaving a
carbon/oxygen core.  The spectrum itself, though derived from the same
underlying mechanism, can be very different depending on the amount of
the envelope stripped from the star at the time of the explosion.  
If the GRB mechanism does entail a jet (e.g., MacFadyen, Woosley, \&
Heger 2001), then a smaller progenitor with a less massive envelope
would make it easier for the jet to punch through the stellar
atmosphere and still have the energy to produce the observed burst and
afterglow.

\section{Other Peculiar Type Ic SNe}

In addition to SN~1998bw, there are several low-redshift peculiar Type
Ic SNe.  These are objects that show high expansion velocities and
sometimes, but not always, large luminosity (they are occasionally
referred to as ``hypernovae'').  Two well-studied such examples are
SN~1997ef and SN~2002ap (Iwamoto et al. 1998, 2000; Foley
et~al. 2003).  None of these other high-velocity SNe has been directly
associated with a GRB.  There is a wide diversity in their
spectroscopic and photometric development, but only the ones like
SN~1998bw have been related to GRBs.  It is not yet clear if this is
significant.

Even though the overall energy budget may be different for GRBs and
these other peculiar Type Ic SNe, they are clearly unusual objects.
The high velocities set them apart from a typical SN, with the only
similar objects being the SNe associated with GRBs.  To study the SN
component of a GRB requires it to be at low redshift, a condition that
also makes them rare.  The peculiar Ic SNe, though, are relatively
more common, and are bright enough to enable detailed studies with
even modest telescopes.  Study of these objects could be the first
step to understanding the GRB mechanism.

\section{Conclusions}
While circumstantial arguments had linked SNe with GRBs, spectroscopic
data was needed for irrefutable proof.  This crucial piece of
evidence arrived in the form of GRB~030329, wherein a direct
transformation from GRB afterglow to peculiar Type Ic SNe was
thoroughly observed.  Later that same year, GRB~031203 provided a
further link.  These two, along with SN~1998bw, showed similar
spectra.  That they were Type Ic implied that the progenitor was a
stripped-envelope star, lending credence to jet-based models for
GRBs.  Although these GRB-SNe are rare, low-redshift examples of
peculiar Type Ic SNe are not, providing a laboratory to understand
stellar explosions in extreme circumstances.

\acknowledgments{I wish to thank my colleagues, Kris Stanek and Peter
Garnavich, for their support and enthusiasm for our group effort
studying the GRB-SN connection.}

\end{document}